\documentclass[]{spie}  

\usepackage{amsmath,amsfonts,amssymb}
\usepackage{graphicx}
\usepackage[colorlinks=true, allcolors=blue]{hyperref}
\usepackage{soul}
\usepackage{afterpage}

\graphicspath{{figures/}}

\title{The conceptual design of GMagAO-X: visible wavelength high contrast imaging with GMT}

\author[a]{Jared R. Males}
\author[a]{Laird M. Close}
\author[a]{Sebastiaan Y. Haffert}
\author[a]{Olivier Guyon}
\author[a]{Victor Gasho}
\author[a]{Fernando Coronado}
\author[a]{Olivier Durney}
\author[b]{Alexander Hedglen}
\author[b]{Maggie Kautz}
\author[a]{Jamison Noenickx}
\author[a]{John Ford}
\author[a]{Tom Connors}
\author[a]{Doug Kelly}
\affil[a]{Steward Observatory, University of Arizona}
\affil[b]{James C. Wyant College of Optical Sciences, University of Arizona}

\authorinfo{Further author information: (Send correspondence to J.R.M.)\\J.R.M.: E-mail: jrmales@arizona.edu}

\begin{document} 
\maketitle

\begin{abstract}
   We present the conceptual design of GMagAO-X, an extreme adaptive optics system for the 25 m Giant Magellan Telescope (GMT).  We are developing GMagAO-X to be available at or shortly after first-light of the GMT, to enable early high contrast exoplanet science in response to the Astro2020 recommendations.  A key science goal is the characterization of nearby potentially habitable terrestrial worlds.  GMagAO-Xis a woofer-tweeter system, with integrated segment phasing control.  The tweeter is a 21,000 actuator segmented deformable mirror, composed of seven 3000 actuator segments.  A multi-stage wavefront sensing system provides for bootstrapping, phasing, and high order sensing.  The entire instrument is mounted in a rotator to provide gravity invariance. After the main AO system, visible (g to y) and near-IR (Y to H) science channels contain integrated coronagraphic wavefront control systems.  The fully corrected and, optionally, coronagraphically filtered beams will then be fed to a suite of focal plane instrumentation including imagers and spectrographs.  This will include existing facility instruments at GMT via fiber feeds.  To assess the design we have developed an end-to-end frequency-domain modeling framework for assessing the performance of GMagAO-X.  The dynamics of the many closed-loop feedback control systems are then modeled.  Finally, we employ a frequency-domain model of post-processing algorithms to analyze the final post-processed sensitivity.  The CoDR for GMagAO-X was held in September, 2021. Here we present an overview of the science cases, instrument design, expected performance, and concept of operations for GMagAO-X.
\end{abstract}


\section{INTRODUCTION}
\label{sec:intro}  
The coming generation of 25-40 m Extremely Large Telescopes (ELTs) promises to transform the study of extrasolar planets.  The improved sensitivity due to larger collection areas compared to today's 5-10 m class telescopes will enable highly sensitive radial velocity searches, as well as transit spectroscopy on a range of nearby planets.  Perhaps the area of exoplanet science which stands to benefit the most is direct imaging.  With adaptive optics  (AO) delivering diffraction limited resolution, the improvement in sensitivity scales as diameter $D^4$.  Just as importantly, the improvement in spatial resolution with $\propto 1/D$ will allow direct imaging at smaller separations than currently possible with today's telescopes.  If we equip them with ``extreme'' AO (ExAO\cite{2018ARA&A..56..315G}) systems and coronagraphs, the ELTs will be able to characterize large numbers of exoplanets.  Most excitingly, such systems will enable us to characterize large numbers of \textit{temperate}, \textit{mature} exoplanets for the first time\cite{2012SPIE.8447E..1XG,2014SPIE.9148E..20M}.  Basic characterization (e.g. broadband albedo measurement, molecule detection) of a few nearby exoplanets with such characteristics may be possible with current telescopes in reflected visible and near-IR light \cite{2017A&A...599A..16L} and the thermal infrared\cite{2021NatCo..12..922W}.  But it is the ELTs which will finally enable detailed studies of solar system like planets.

The most profound result of this transformative leap in capabilities will be the ability to search for the signatures of life on planets orbiting other stars.  The Astro2020 Decadal Survey\cite{2021pdaa.book.....N} established ``Pathways to Habitable Worlds''  placing the search for life on planets other than Earth as a Priority Area for astrophysics over the next decade.  One of the four key capabilities needed to achieve this is: \textbf{\textit{Ground-based extremely large telescopes equipped with high-resolution spectroscopy, high-performance adaptive optics, and high-contrast imaging.}}\cite{2021pdaa.book.....N}

The Giant Magellan Telescope (GMT) presents a unique opportunity to realize the vision of Astro2020.  With the smallest diameter of the 3 planned ELTs, it can achieve higher Strehl with fewer deformable mirror (DM) actuators -- making it an ideal facility for short wavelength ExAO \cite{2014SPIE.9148E..1MC}.  Furthermore, the large-segment design of the GMT provides a straightforward way to harness existing technologies to assemble the required DM.  Finally, the current instrument suite of the GMT\cite{Sitarski_2022} has space for an ExAO instrument, which could then be coupled to the already in-progress high resolution spectrographs at GMT.  Here we present the results of the conceptual design study for GMagAO-X, an instrument in development to fill this niche and be ready at or shortly after first light of the GMT. At the conclusion of the conceptual design study, GMagAO-X underwent a conceptual design review (CoDR) in September, 2021, with an external panel, and is now in the preliminary design phase.

\section{SCIENCE OBJECTIVES}
\label{sec:science}

The diffraction limited resolution resolution of the GMT at 500 nm (V band) will be 4.1 mas, and at 1600 nm (H band) it will be 13.2 mas.  These unprecedented spatial resolutions, when coupled with $D^4$ exposure time scaling, will make the GMT a  truly transformative scientific instrument.  Taking advantage of this, GMagAO-X will enable high angular resolution investigations of many compelling science cases.  Here we list the top priority science objectives identified for CoDR, all of which address the Astro2020 theme of ``Worlds and Suns in Context'' which ``...encompasses the interlinked studies of stars, planetary systems, and the solar system.''\cite{2021pdaa.book.....N}.
\begin{enumerate}

\item \textbf{Search For Life} [new science]
   \begin{itemize}
      \item Search for life on terrestrial exoplanets orbiting nearby stars
   \end{itemize}

\item \textbf{Characterize Older Temperate Exoplanets} [new science]
   \begin{itemize}
      \item Reflected light characterization of wide range of radii and mass
   \end{itemize}

\item \textbf{Measure Orbit and Mass of Exoplanets} [today’s science, better]
   \begin{itemize}
      \item Feed to G-CLEF for Precision-RV 
      \item Identify new candidates
      \item Obtain precise ephemerides for imaging targets
   \end{itemize}

\item \textbf{Study Planet Formation at Low Mass and Small Separation} [today’s science, better]
   \begin{itemize}
      \item Search for and characterize thermally self-luminous young planets (YJH)
      \item Characterize forming planets through the H-alpha accretion signature
   \end{itemize}

\item \textbf{Circumstellar Disk Structure and Disk-Planet Interactions} [today’s science, better]
   \begin{itemize}
      \item Structure in scattered light at high spatial resolution
      \item Study disk-planet interactions
      \item Jets and outflows
   \end{itemize}

\item \textbf{Refine Stellar Evolution Models} [today’s science, better]
   \begin{itemize}
      \item High spatial and spectral resolution characterization of benchmark binaries
   \end{itemize}

\item \textbf{High Spatial Resolution Mapping} [today’s science, better]
   \begin{itemize}
      \item Moons, asteroids, stellar surfaces
   \end{itemize}

\end{enumerate}

The key rationale for the conceptual design of GMagAO-X  is that the first two science objectives, which require reflected light characterization of temperate exoplanets, are the most demanding.  An ExAO instrument capable of meeting these top science objectives will therefore be capable of meeting the others.  Placing our emphasis on these objectives is validated by Astro2020's ``Pathways to Habitable Worlds''.

\section{INSTRUMENT OVERVIEW}

\subsection{Optical and Mechanical Design}

 The optical and mechanical design of GMagAO-X is presented in detail in these proceedings\cite{Close_2022}, as are the details of our wavefront sensing strategy\cite{Haffert_2022}, and the plans for segment phasing control\cite{Hedglen_2022} and the 21,000 actuator ``parallel DM''\cite{Kautz_2022}.  In brief, the main components of GMagAO-X consist of
\begin{itemize}
   \item Woofer DM: ALPAO DM 3228 (64 actuator diameter) \cite{Close_2022}
   \item 1st Stage WFS: 64x64 infrared Pyramid WFS \cite{Haffert_2022}
   \item Phase Control: 7x piezo T/T and Piston stages \cite{Close_2022,Hedglen_2022}
   \item Tweeter DM: seven 3000 actuator MEMS DMs\cite{Close_2022,Kautz_2022}
   \item Coarse Phase Sensing: Holographic Dispersed Fringe Sensor\cite{10.1117/1.JATIS.8.2.021513,10.1117/1.JATIS.8.2.021515} (HDFS, probably IR using 10\% of J-H)
   \item 2nd Stage WFS \& Fine Phase Sensing: 212x212 PWFS (I-J bands).  Also considering Zernike WFS  \cite{Haffert_2022}
   \item Coronagraph NCP DM: 3000 actuator MEMS DM
   \item Coronagraph: Lyot-architecture, supporting PIAACMC
   \item Science: SDI with dual cameras, fiber feeds to other GMT instruments (namely G-CLEF)
\end{itemize}

GMagAO-X is designed to be build-able today, with existing technology.  All of the above components can be purchased or fabricated today, making this instrument well suited to the early-science phase at GMT.

\begin{figure}[h]
   \centering
   \includegraphics[width=4.5in]{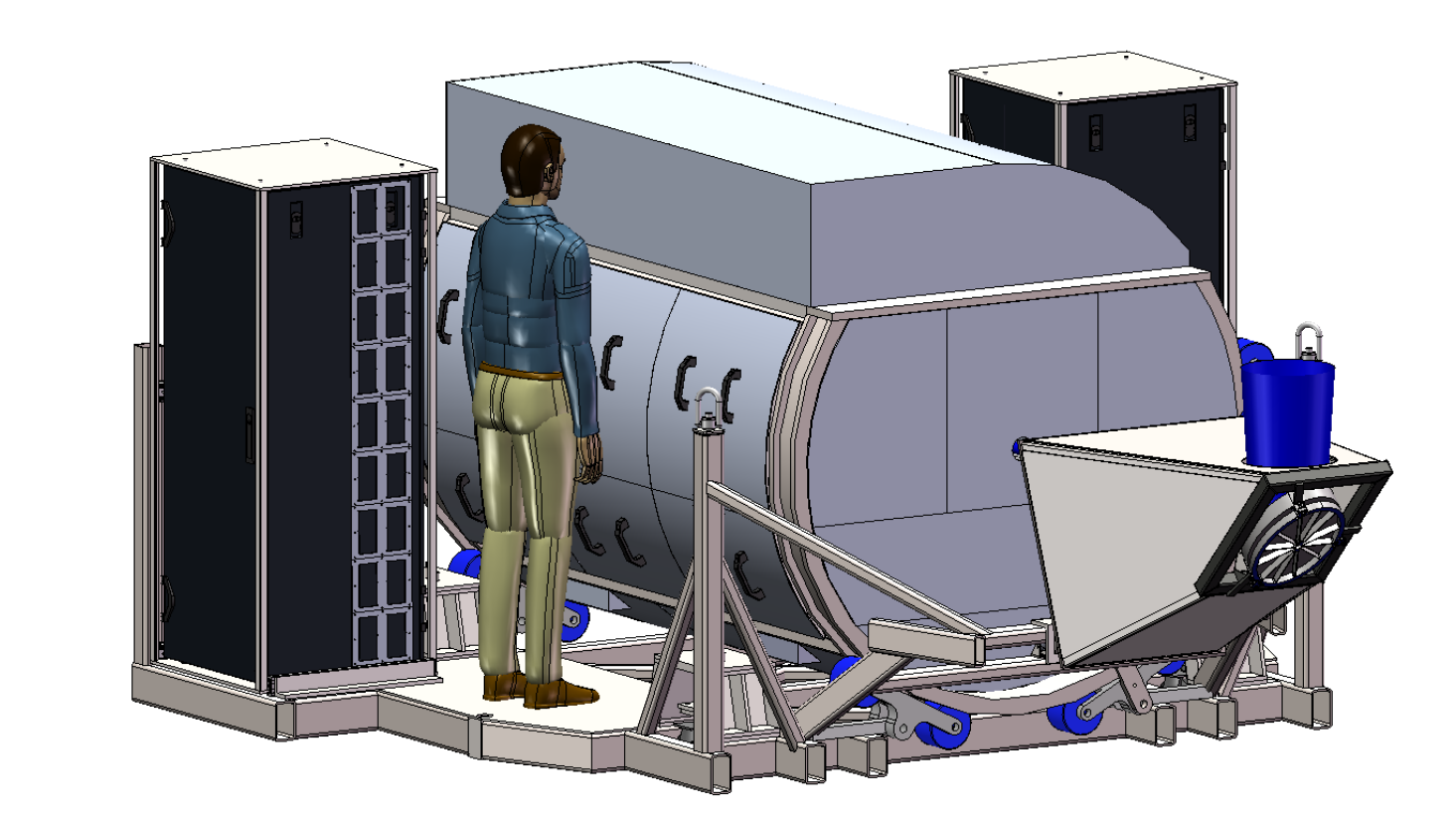}
   \caption{ Iso view of GMagAO-X.  See text for details. \label{fig:instrument_main_view}}
\end{figure}

Figure \ref{fig:instrument_main_view} shows an iso view of GMagAO-X, which will be an air-isolated two-level floating optical table mounted on the Gregorian instrument rotator (GIR) in a folded port (FP).  The optical table system will be housed inside a barrel which rotates to maintain gravity down as the GMT elevation changes during an observation.  We plan to use a deployable tertiary mirror, rather than the facility M3, to allow much higher optical quality on a smaller optic.  See Close et al in these proceedings for a more complete description.\cite{Close_2022}

\subsection{Electronics}

Shown in Figure \ref{fig:instrument_main_view}, non-rotating electronics racks house control equipment for the rotator, as well as the AO and coronagraph components.  Deformable mirror controllers will consume one of the racks shown.  Computing equipment will be housed off the telescope mount with low-latency fiber communications in between.

The on-instrument electronic devices, such as filter wheels and focus stages, are based on similar devices in MagAO-X\cite{2018SPIE10703E..09M,2020SPIE11448E..4LM,Males_2022}.  

\subsection{Software}

The instrument rotation control software, as well as safety and interlock systems, will be based on standard GMT components.  This will ensure that GMagAO-X can be kept safe during any telescope operation when not observing.

Instrument control software will be based on the now well-tested MagAO-X instrument control software.  This includes the CACAO real-time software system.\cite{2018SPIE10703E..1EG}

\section{AO ERROR BUDGET}

\begin{figure}[h]
   \centering
   \includegraphics[width=\textwidth]{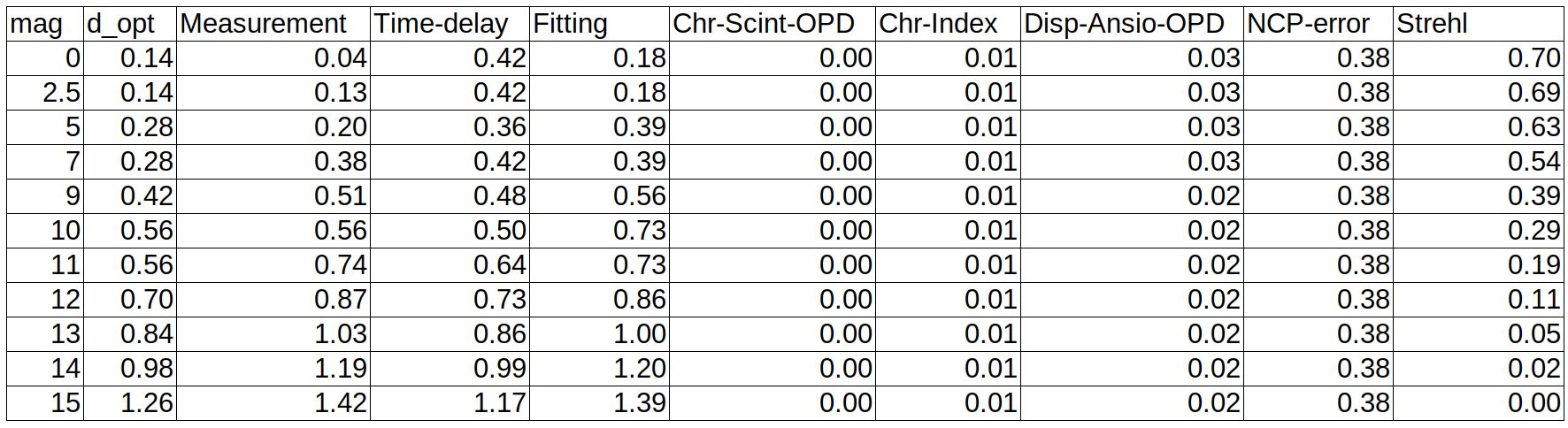}
   \caption{Simple error budget for GMagAO-X for WFS operating at 800 nm and observations at 850 nm. \label{fig:err_budg_850_800}}
\end{figure}

Based on the chosen AO components and parameters, we first calculated a simple error budget.  The results are shown in Figure \ref{fig:err_budg_850_800} for the WFS operating at 800 nm and science observations at 850 nm.  For each guide star magnitude, the d\_opt column shows the optimum actuator spacing, which is an integer multiple of 14 cm.  This is to account for WFS detector binning.  WFS exposure time is optimized at each magnitude.  The resulting wavefront error terms are shown as a function of guide star magnitude in units of radians at 800 nm.  Included are measurement (WFS noise), time-delay, fitting (a function of d\_opt), chromaticity of scintillation affecting the OPD, chromaticity of the index of refraction, dispersive anisoplanatism, and static non-common path errors.  The resulting Strehl ratio is shown in the last column.

\begin{figure}[h]
   \centering
   \includegraphics[width=\textwidth]{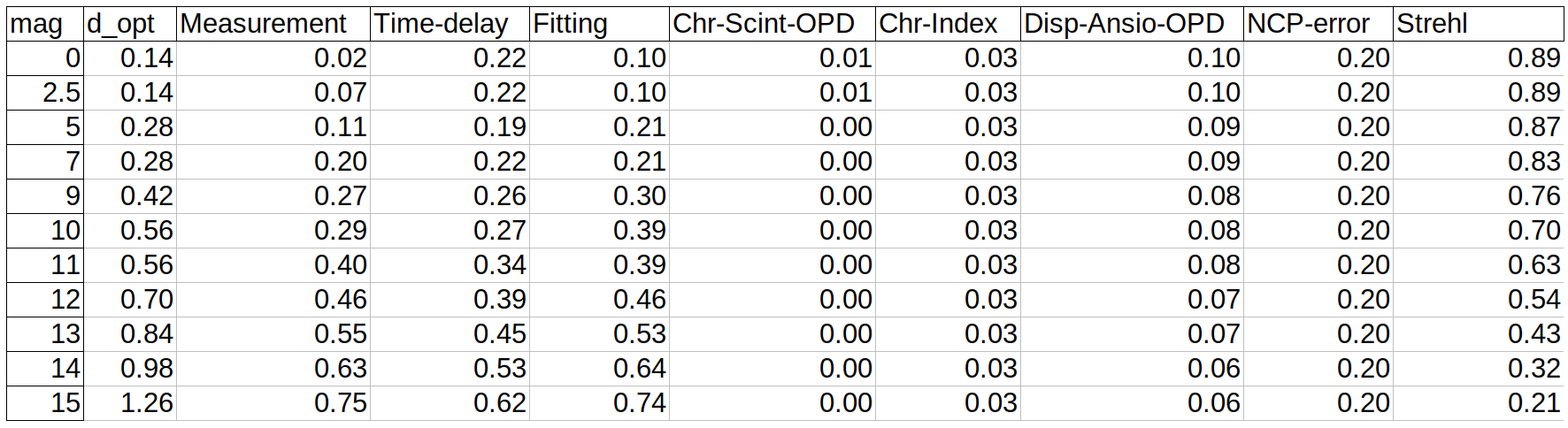}
   \caption{Simple error budget for GMagAO-X for WFS operating at 800 nm and observations at 1600 nm. \label{fig:err_budg_1600_800}}
\end{figure}

Figure \ref{fig:err_budg_1600_800} is the same except for a science observing wavelength of 1600 nm.

\section{E2E PERFORMANCE MODELING}

\begin{figure}[h]
   \centering
   \includegraphics[width=\textwidth]{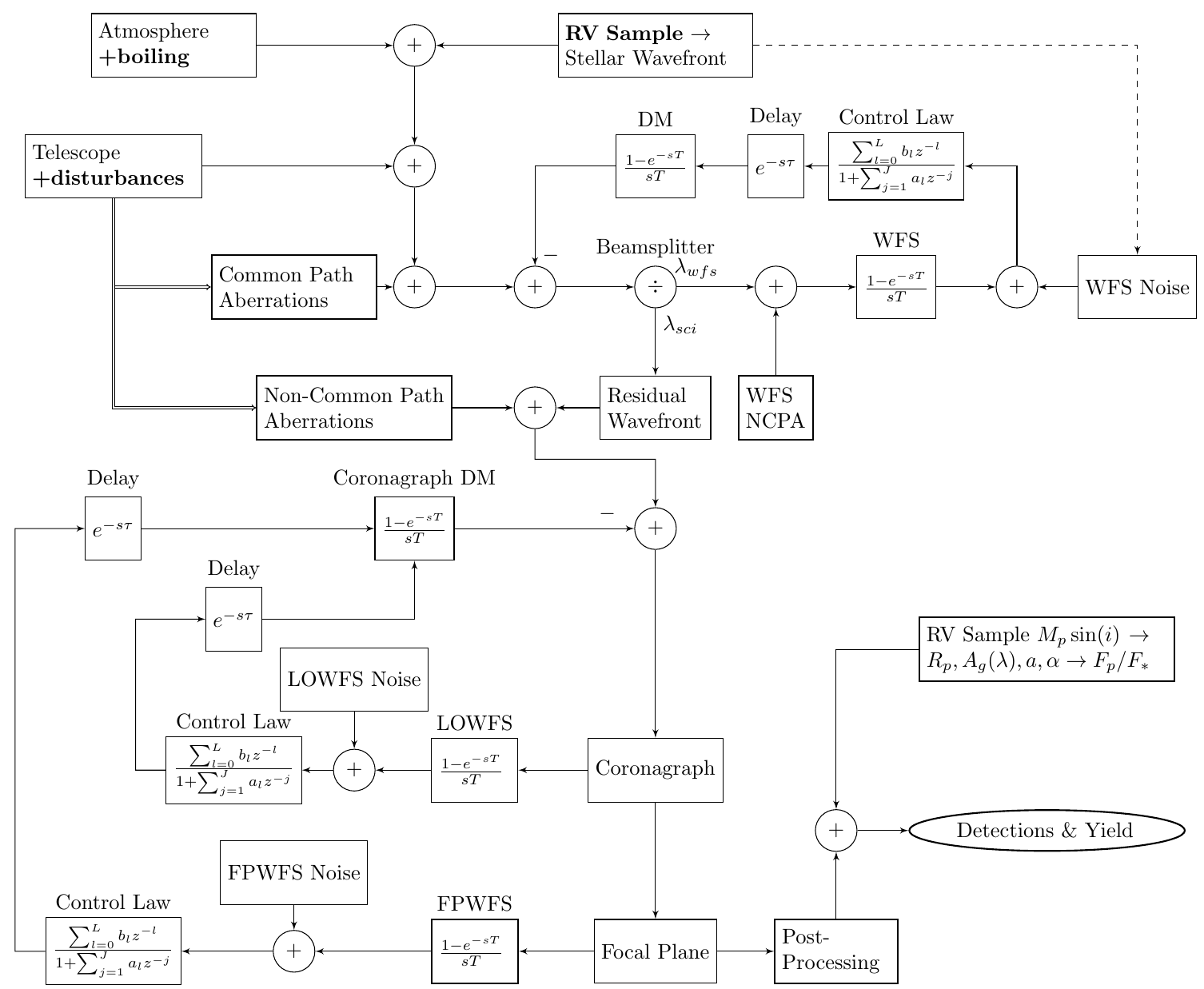}
   \caption{Illustration of the end-to-end semi-analytic model under development for GMagAO-X.  The model treats control system dynamics and produces the post-coronagraph contrast\cite{2018JATIS...4a9001M} and intensity power spectrum\cite{2021PASP..133j4504M} at any location in the focal plane. \label{fig:e2e_model}}
\end{figure}

We next developed an end-to-end (E2E) model of GMagAO-X.  We have been developing a semi-analytic frequency domain framework for such analyses which models coronagraph performance\cite{2018JATIS...4a9001M}, and we have extended it to include focal plane intensity dynamics and post-processing \cite{2021PASP..133j4504M}.  The model explicitly includes modal gain optimization, and for this study we treated only frozen-flow turbulence and the high-order AO loop.  Figure \ref{fig:e2e_model} illustrates the E2E model.

\begin{figure}[h]
   \centering
   \includegraphics[width=3in]{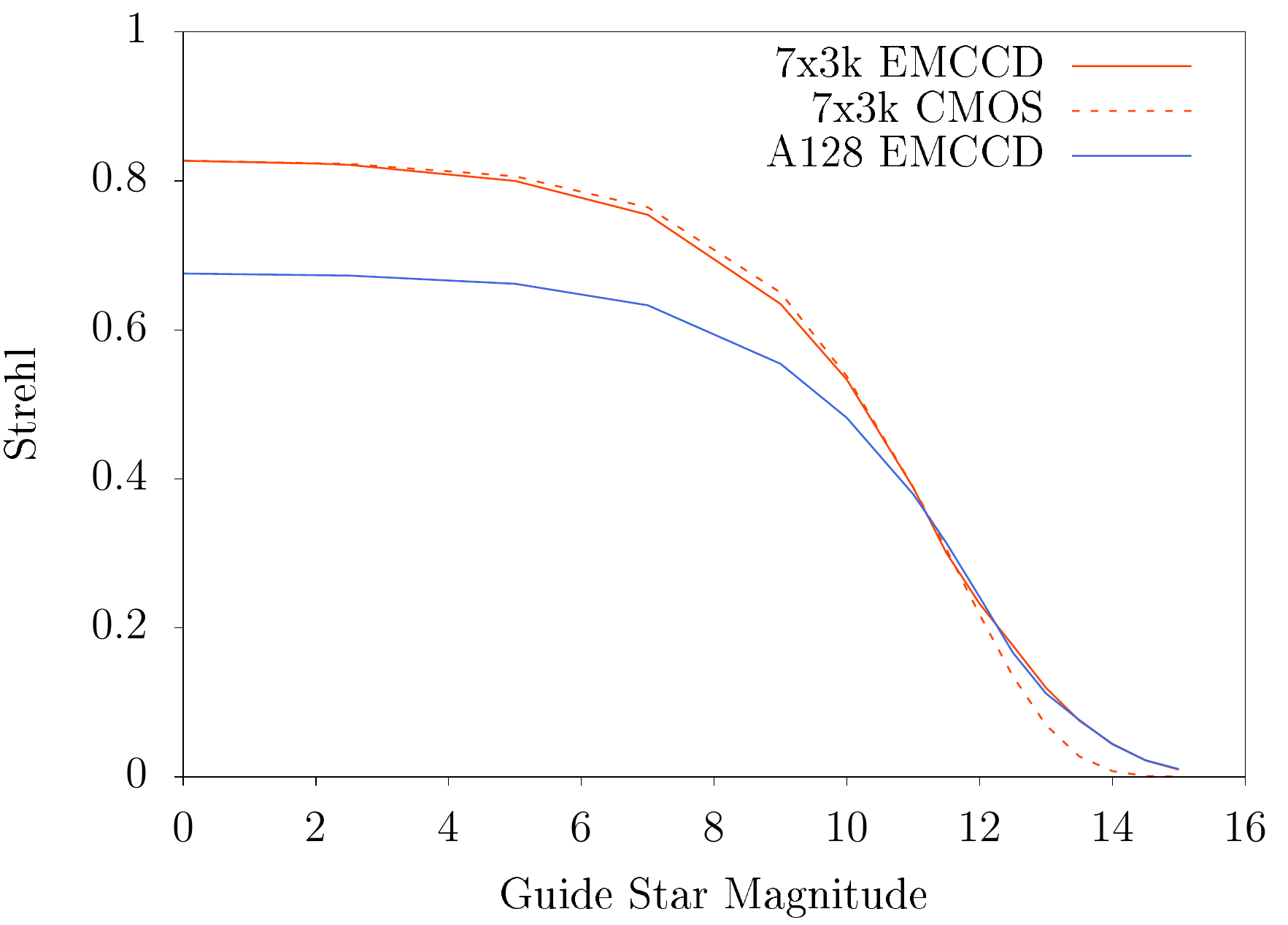}
   \caption{Comparison between the 7x3k ``Parallel DM'' and a monolithic 128x128 actuator DM in terms of Strehl ratio at 800 nm.  The fitting error term is significant for the largest currently planned monolithic DM.  Such a DM with roughly 190 actuators across will be required to match the Parallel DM, and such devices may be possible in the future but we do not expect them to be available on the timescales of GMagAO-X.  \label{fig:strehl_dm_trade}}
\end{figure}

We used the E2E model to perform a trade study to compare our chosen 7x3k ``parallel DM''\cite{Close_2022,Kautz_2022} to a plausible monolithic DM with 128x128 actuators.  The 7x3k has 190 effective actuators across.  As shown in Figure \ref{fig:strehl_dm_trade} the fitting error at 800 nm makes the 128x128 monolithic DM a secondary choice for GMagAO-X. 

We also compared using an EMCCD to a CMOS detector.  The EMCCD was assigned readout noise of 0.5 electrons, and the CMOS 2 electrons.  The EMCCD excess noise factor was included.  The CMOS WFS performs better on $\sim4$ to $\sim10$ magnitude stars where the excess noise is important.  The lower readnoise of the EMCCD provides better performance on fainter stars.

\begin{figure}[h!]
   \centering
   \includegraphics[width=4.5in]{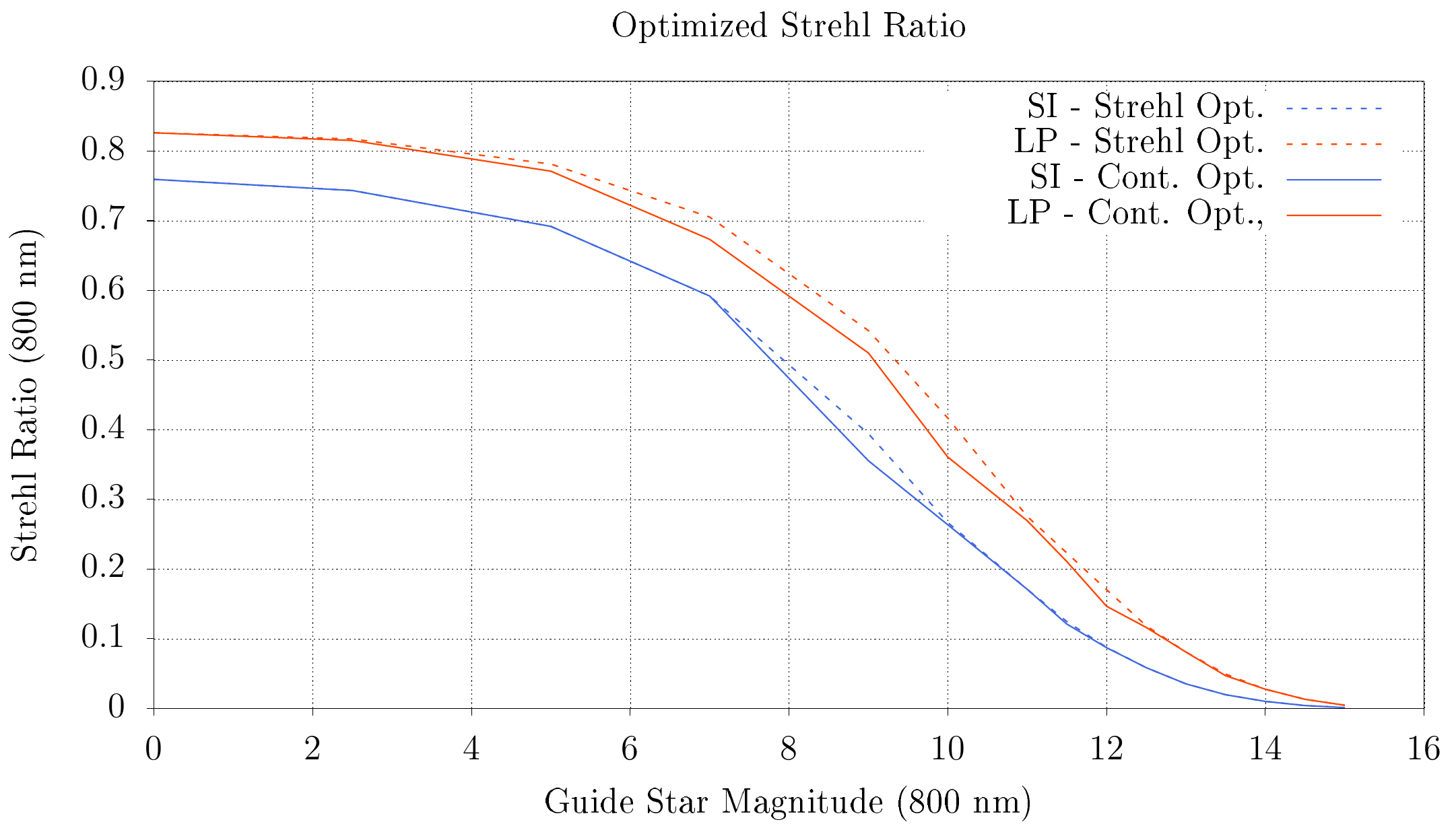}
   \caption{Predicted Strehl ratio when the end-to-end model of GMagAO-X is optimized for best Strehl ratio and for best contrast at 3 $\lambda/D$.  The linear predictor (LP) controller has a significant impact over the simple integrator (SI). \label{fig:strehl_optimized}}
\end{figure}

\begin{figure}[h!]
   \centering
   \includegraphics[width=4.5in]{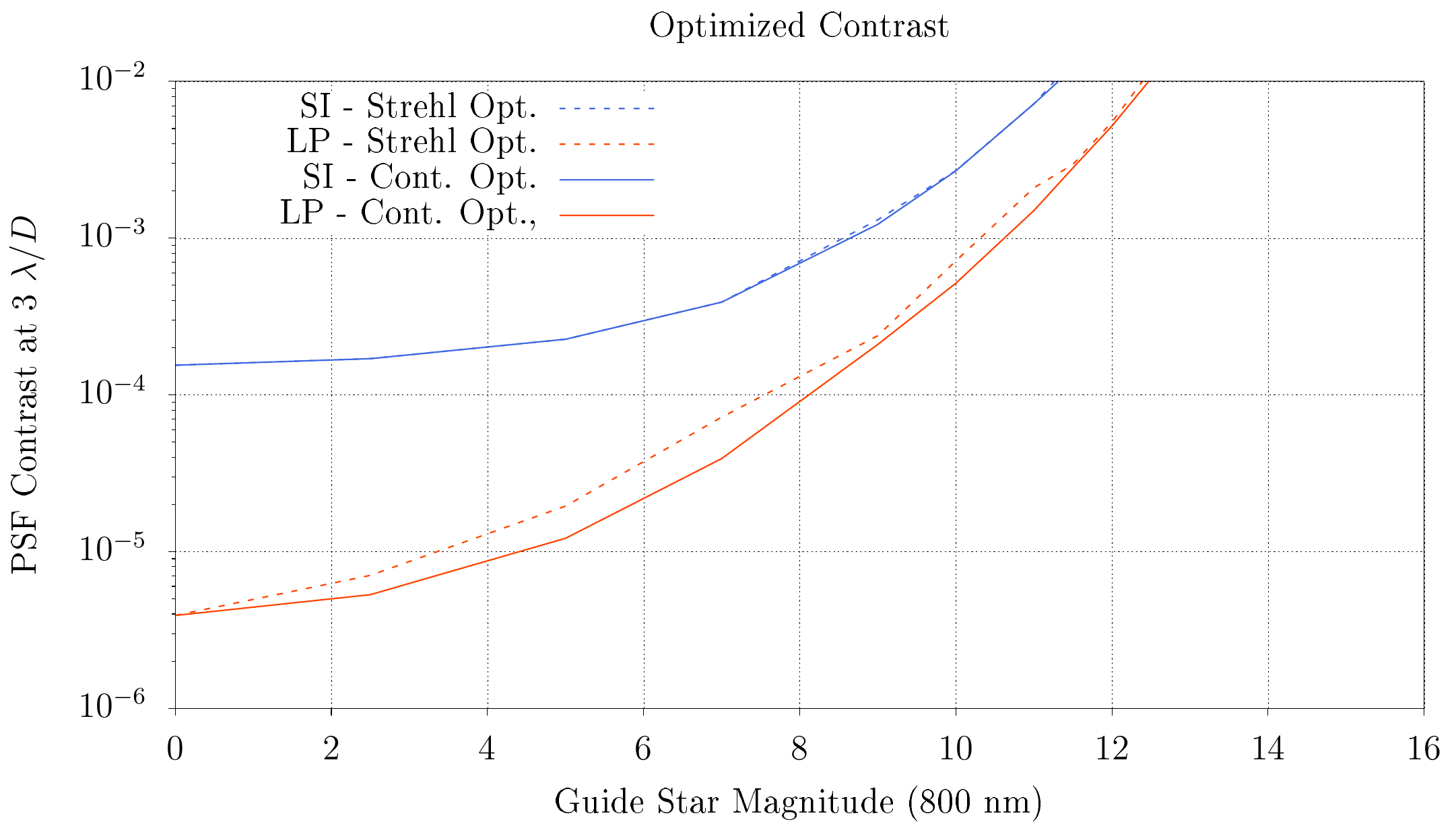}
   \caption{Predicted raw PSF contrast ratio when the end-to-end model of GMagAO-X is optimized for Strehl ratio and for best contrast at 3 $\lambda/D$. The linear predictor (LP) controller has a significant impact over the simple integrator (SI).  \label{fig:contrast_optimized}}
\end{figure}

We next computed the optimum Strehl (Figure \ref{fig:strehl_optimized}) and contrast at 3 $\lambda/D$ (Figure \ref{fig:contrast_optimized}).  Here we compare the simple integrator (SI) and the Linear Predictor (LP)\cite{2018JATIS...4a9001M}.  The results depend modestly on whether Strehl or contrast is used as the criterion for optimization.  The predictive controller shows a significant gain in post-coronagraph contrast as expected.

\section{EXOPLANET YIELD}

To assess the suitability of the conceptual design for meeting the science objectives, we next analyzed the temperate exoplanet yield of GMagAO-X.  This was based on the catalog of known exoplanets maintained by NExScI.  The catalog was queried to identify exoplanet host stars within the declination range of GMT.  The parameters of these stars were first normalized using standard main sequence relationships, filling in any missing values such as luminosity or colors.  Next, the exoplanets were analyzed for detectability.  Minimum mass was converted to mass assuming $\sin(i)=60^o$, and this was then converted to radius using a mass-to-radius relationship we developed for use on these lightly-irradiated planets.  The Lambertian phase function was assumed.  We consider a range of models for geometric albedo $A_g$.  For the results presented here, we assume a constant  $A_g=0.4$, which is more conservative than the mean value of models such as the Earth, Venus, and gas-giant models such as those of\cite{2010ApJ...724..189C}.    The planet:star flux ratio of each planet as a function of its orbit was then computed.  The output of the E2E model described above was then converted to a photon noise and a speckle noise contribution, and the signal-to-noise ratio as a function of time was computed.  Here we assume that quasi-static speckles are controlled by the focal-plane and low-order WFS strategies designed into  GMagAO-X sufficiently that the main contributor to the photon and speckle noise terms is residual atmospheric turbulence.  We initially assume a coronagraph with an inner working angle (IWA) of $1\lambda/D$, an assumption we will later relax.

\begin{figure}[h]
   \centering
   \includegraphics[width=4.0in]{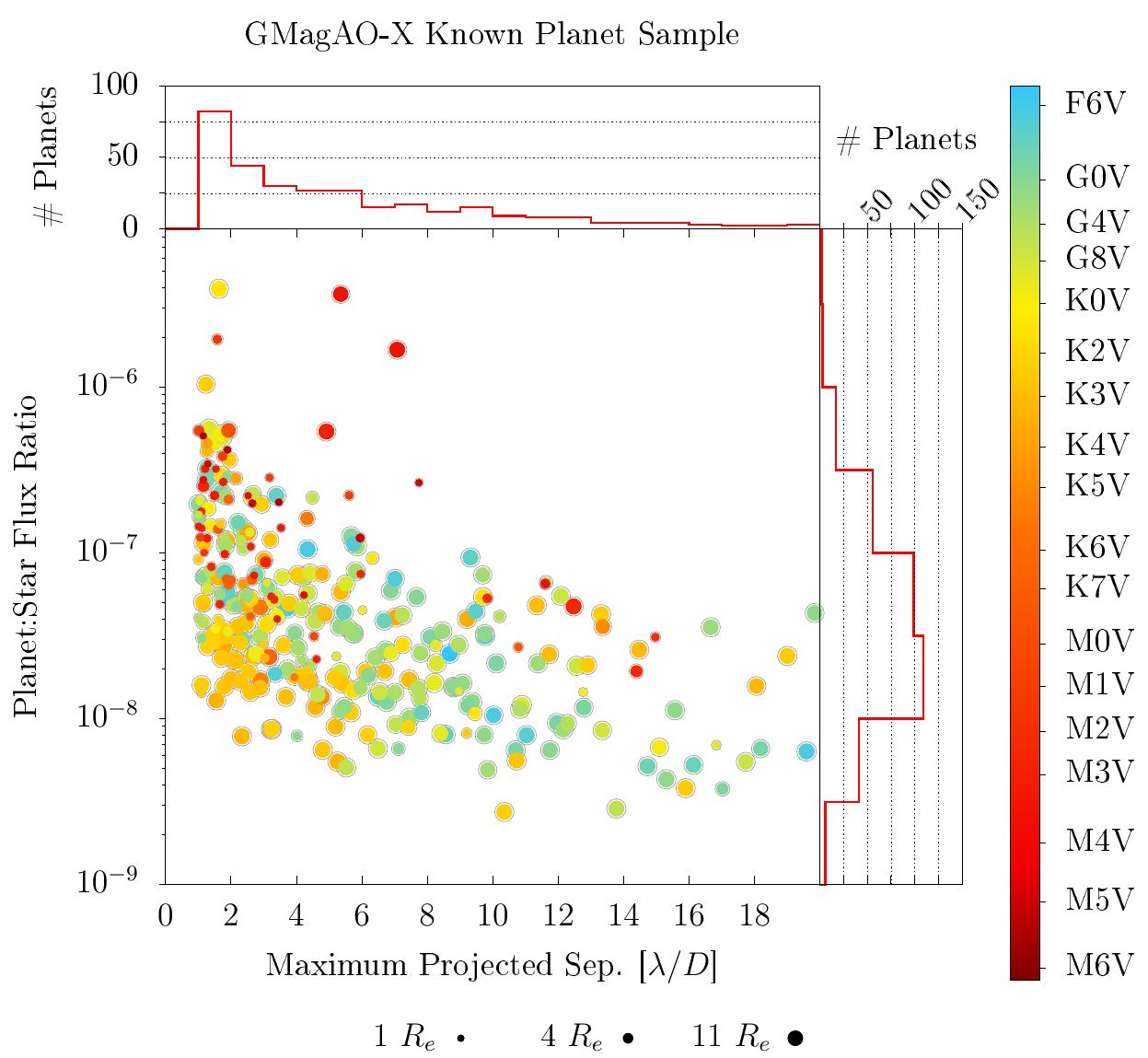}
   \caption{Planet:Star flux ratio vs. separation in $\lambda/D$ at 800 nm for the GMagAO-X RV-detected planet sample.  \label{fig:planets_contrast_v_sep}}
\end{figure}

\begin{figure}[h]
   \centering
   \includegraphics[width=4.0in]{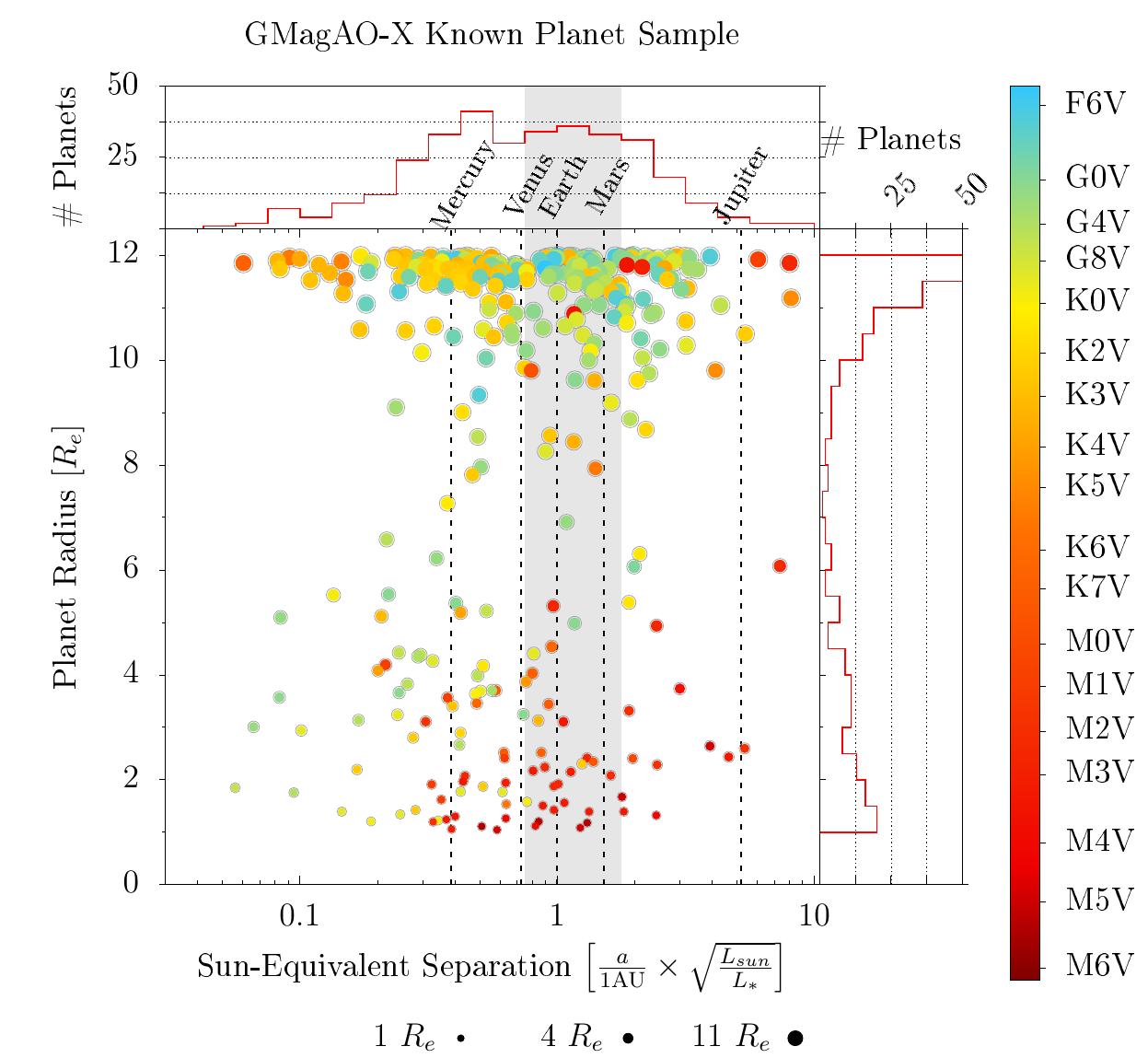}
   \caption{Characteristics of the GMagAO-X RV-detected planet sample.  Against the y-axis we plot planet radius, inferred by assuming $\sin(i) = 60^o$ and using a custom mass-to-radius conversion.  The x-axis shows semi-major axis normalized by the square root of the stellar luminosity.  The gray region denotes the liquid water habitable zone.  \label{fig:planets_rad_v_seff}}
\end{figure}

Figure \ref{fig:planets_contrast_v_sep} displays the planet:star flux ratio vs separation for all of the exoplanets we assess to be detectable by GMagAO-X.  In Figure \ref{fig:planets_rad_v_seff} we show the key physical characteristics of these planets, radius vs. luminosity normalized separation (a proxy for temperature).  Most of the planets have planet:star flux ratios $>10^{-8}$ (meaning they are brighter than this).  The majority are temperate gas-giants, and a range of host spectral types are represented.  There are $\sim$10 currently known terrestrial planets in the liquid water habitable zones of their stars.  It is these planets which we will search for the signatures of life.

\begin{figure}[h]
   \centering
   \includegraphics[width=4.5in]{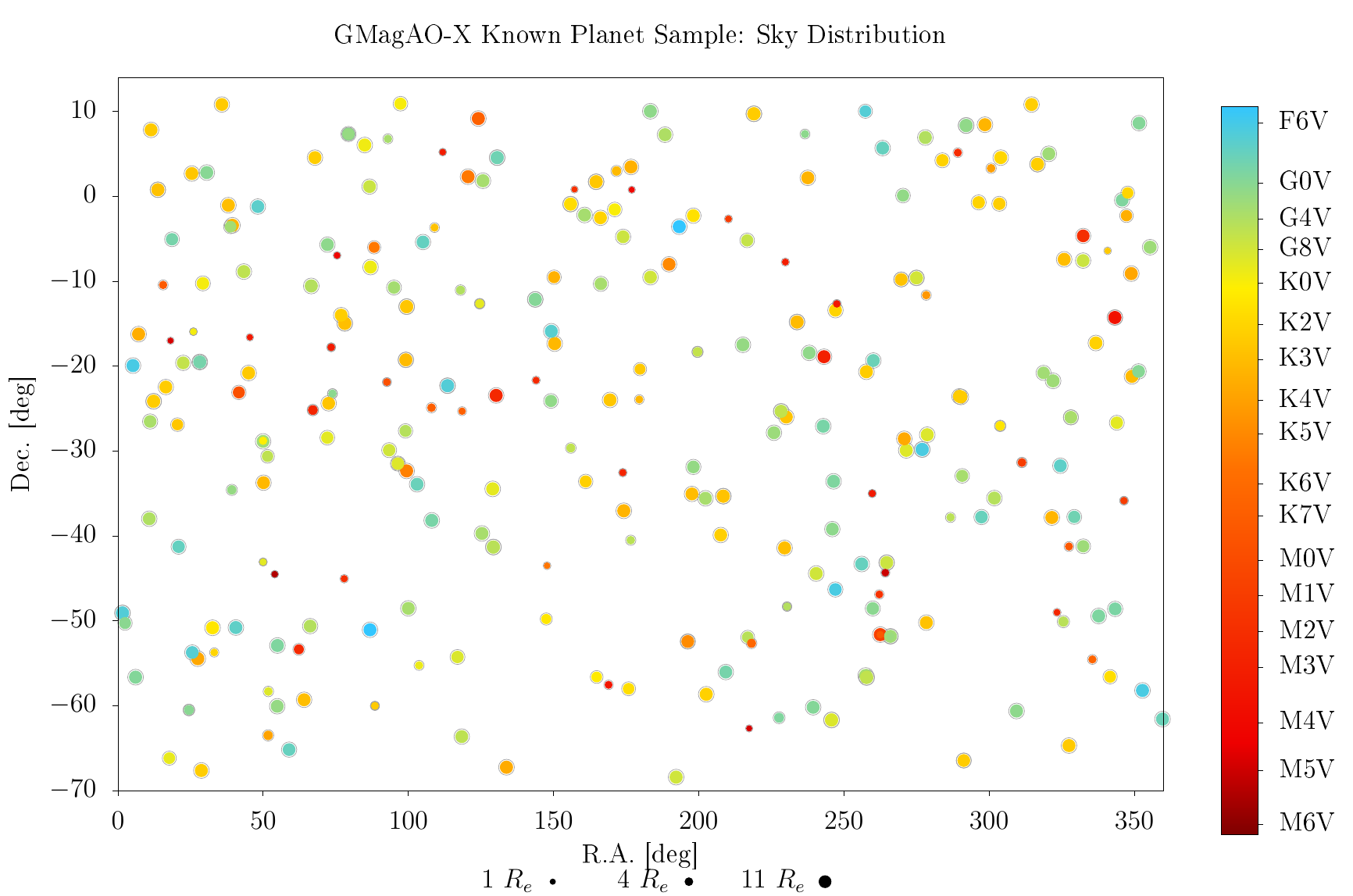}
   \caption{Spatial distribution of the GMagAO-X RV-detected planet sample.  There will be several planets observable on any given night.  \label{fig:planets_ra_dec}}
\end{figure}

In Figure \ref{fig:planets_ra_dec} we show the spatial distribution, in right ascension and declination, of these targets.  This is a key input to the potential concept of operations for an instrument like GMagAO-X, since it motivates a queue mode campaign where anytime conditions are favorable and a known planet hosting star is near transit it can be observed.  This will allow efficiently building exposure time for detailed characterization of these exoplanets.

\begin{figure}[h]
   \centering
   \includegraphics[width=3.0in]{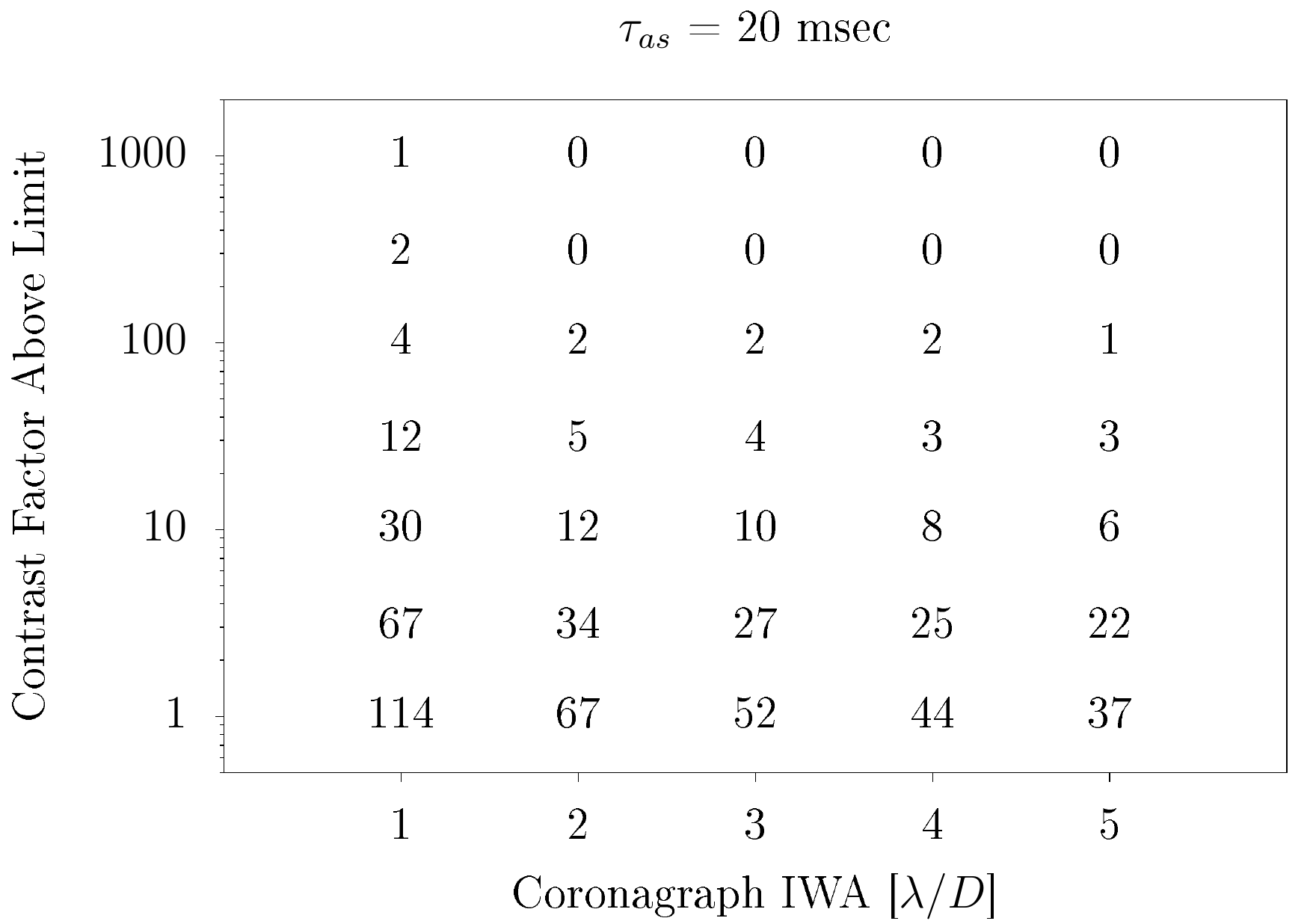}
   \includegraphics[width=3.0in]{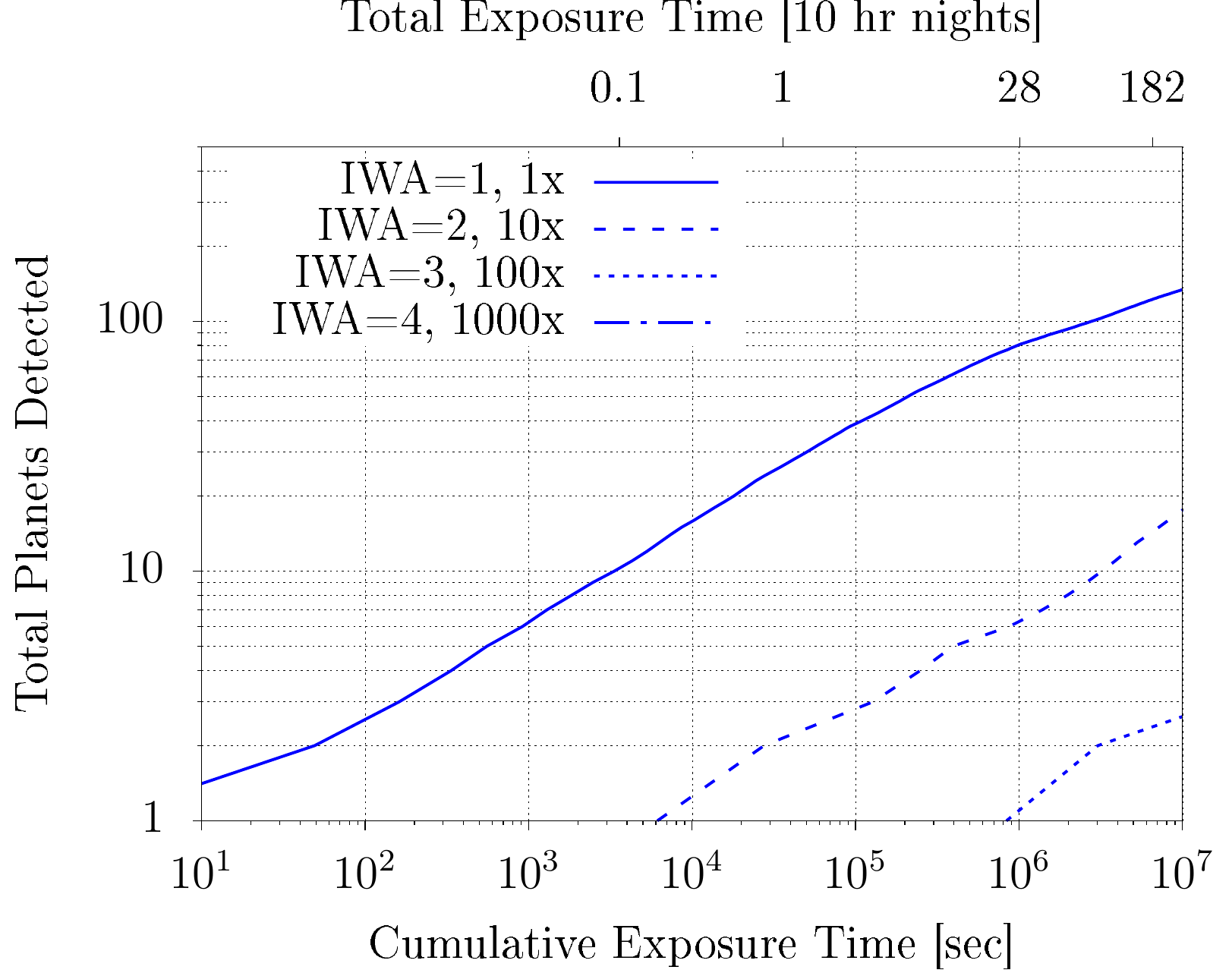}
   \caption{Number of planets characterized with an albedo measurement for various performance degradations assuming a 20 msec residual speckle lifetime.  Left shows the total number at each combination of degradations, and right shown the number of planets vs. total exposure time for the survey.  \label{fig:total_exp_tau20}}
\end{figure}

Finally, we analyze the impact of performance worse than the optimistic assumptions made.  We compared coronagraph IWA of 1, 2, 3, 4, and 5 $\lambda/D$, along with post-coronagraph contrast worse by multiplicative factors of 1, 10, 100, and 1000, respectively.  Figure \ref{fig:total_exp_tau20} shows the impact of these degradations assuming a 20 msec atmospheric speckle lifetime\cite{2021PASP..133j4504M}.   

\begin{figure}[h]
   \centering
   \includegraphics[width=3.0in]{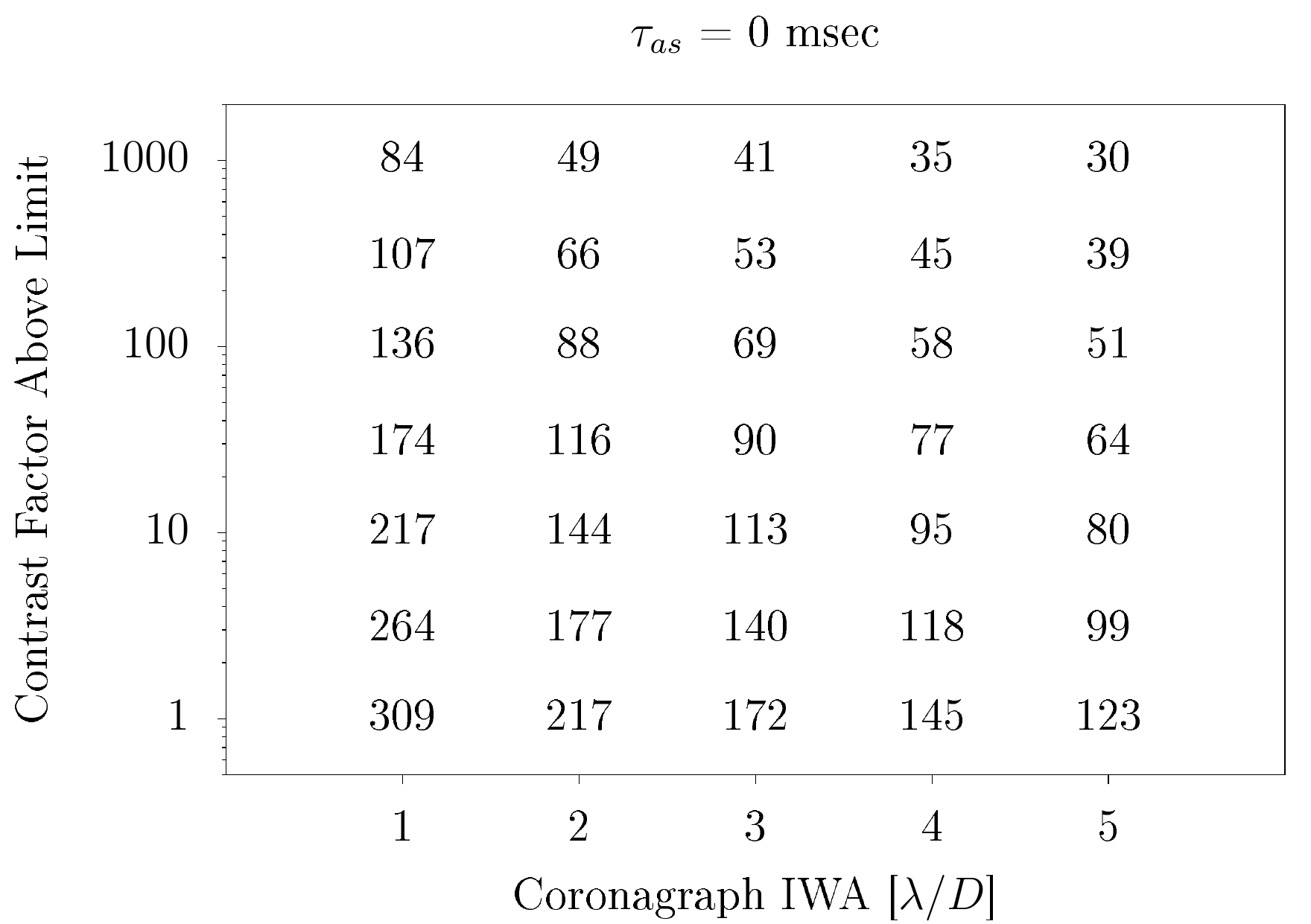}
   \includegraphics[width=3.0in]{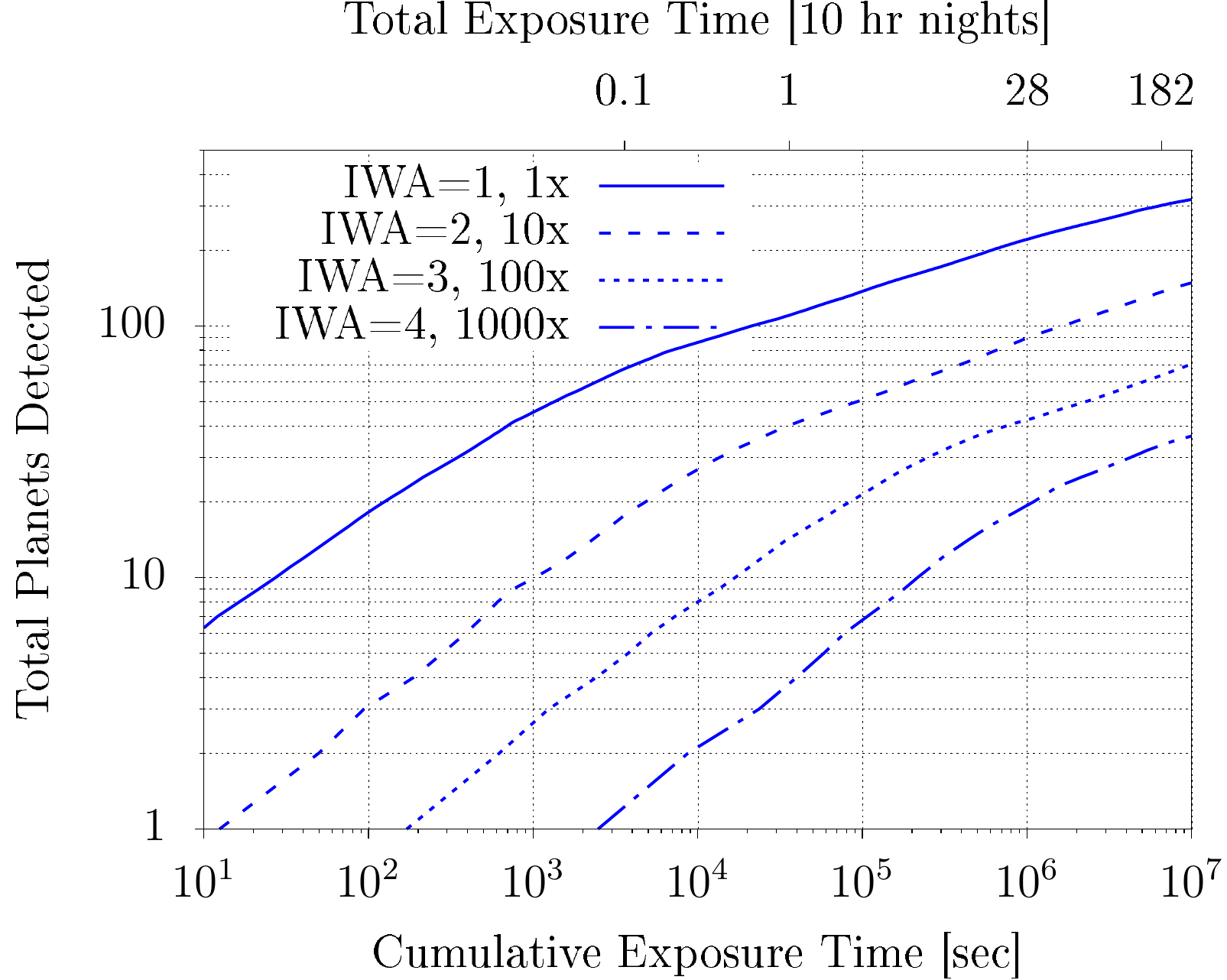}
   \caption{Number of planets characterized with an albedo measurement for various performance degradations assuming a 0 msec \textit{effective} residual speckle lifetime.  Left shows the total number at each combination of degradations, and right shown the number of planets vs. total exposure time for the survey.  We expect to achieve this \textit{effective} residual speckle lifetime through telemetry-based post-processing\cite{2021PASP..133j4504M}.  Compare the dramatic improvement from Figure \ref{fig:total_exp_tau20}.  \label{fig:total_exp_tau0}}
\end{figure}

We also consider post-processing, specifically using WFS telemetry.  This will lower the speckle lifetime sufficiently to make such observations photon-noise dominated, making the \textit{effective} residual speckle lifetime 0\cite{2021PASP..133j4504M}.  See also Guyon et. al. in these proceedings\cite{Guyon_2022}.  The dramatic improvement in yield is illustrated by comparing Figure \ref{fig:total_exp_tau0} to Figure \ref{fig:total_exp_tau20}.

\section{CONCLUSION}

We have completed the conceptual design of GMagAO-X, a 21,000 actuator ExAO coronagraph for the GMT, and successfully completed CoDR in September, 2021.  GMagAO-X takes advantage of GMT's unique design, enabling rapid development and deployment using currently available technologies.  When on-sky at, or shortly after, first-light of the GMT, GMagAO-X will enable the characterization of hundreds of nearby, known-from-RV, temperate, mature extrasolar planets.  Roughly 10 of these are terrestrial planets in the liquid water habitable zone of their host stars.  Such planets are prime targets to search for life to meet the challenge of Astro2020's ``Pathways to Habitable Worlds'' priority area.

\acknowledgments 
 
We are very grateful for support from the University of Arizona Space Institute (UASI).  This research has made use of the NASA Exoplanet Archive, which is operated by the California Institute of Technology, under contract with the National Aeronautics and Space Administration under the Exoplanet Exploration Program.


\bibliography{report} 
\bibliographystyle{spiebib} 

\end{document}